\documentclass[12pt]{article}

\usepackage{scicite}
\usepackage{graphicx}
\usepackage{physics}
\usepackage{multirow}

\usepackage{times}

\newenvironment{sciabstract}{%
\begin{quote} \bf}
{\end{quote}}

\title{Training of Quantum Circuits on a Hybrid Quantum Computer}

\author
{D. Zhu,$^{1\ast}$ N. M. Linke,$^{1}$ M. Benedetti,$^{3,4}$ K. A. Landsman$^{1}$, N. H. Nguyen$^{1}$,\\ C. H. Alderete$^{1}$, A. Perdomo-Ortiz$^{3,7}$, N. Korda$^{5}$, A. Garfoot$^{5}$,\\ C. Brecque$^{5}$, L. Egan$^{1}$, O. Perdomo$^{6}$, C. Monroe$^{1,2}$\\
\normalsize{$^{1}$Joint Quantum Institute, Department of Physics,}\\
\normalsize{and Joint Center for Quantum Information and Computer Science,}\\
\normalsize{University of Maryland, College Park, MD 20742, USA}\\ 
\normalsize{$^{2}$IonQ, Inc., College Park, MD 20740}\\
\normalsize{$^{3}$Department of Computer Science, University College London,}\\
\normalsize{WC1E 6BT London, UK,}\\
\normalsize{$^{4}$Cambridge Quantum Computing Limited, CB2 1UB Cambridge, UK}\\
\normalsize{$^{5}$Mind Foundry Limited, OX2 7DD Oxford, UK}\\
\normalsize{$^{6}$Department of Mathematics, Central Connecticut State University,}\\
\normalsize{New Britain, CT 06050, USA}\\
\normalsize{$^{7}$Zapata Computing Inc., 439 University Avenue,}\\
\normalsize{Office 535, Toronto, ON, M5G 1Y8}\\
\normalsize{$^\ast$To whom correspondence should be addressed; E-mail:  daiwei@terpmail.umd.edu.}
}

\date{}

\begin{document} 


\baselineskip24pt


\maketitle


\begin{sciabstract}
Generative modeling is a flavor of machine learning with applications ranging from computer vision to chemical design. It is expected to be one of the techniques most suited to take advantage of the additional resources provided by near-term quantum computers. Here we implement a data-driven quantum circuit training algorithm on the canonical Bars-and-Stripes data set using a quantum-classical hybrid machine. The training proceeds by running parameterized circuits on a trapped ion quantum computer, and feeding the results to a classical optimizer. We apply two separate strategies, Particle Swarm and Bayesian optimization to this task. We show that the convergence of the quantum circuit to the target distribution depends critically on both the quantum hardware and classical optimization strategy. Our study represents the first successful training of a high-dimensional universal quantum circuit, and highlights the promise and challenges associated with hybrid learning schemes.
\end{sciabstract}

\section*{One Sentence Summary}
We train generative modeling circuits on a quantum-classical hybrid computer showing optimization strategy and resource trade-off.

\section*{Introduction}

Hybrid quantum algorithms \cite{mcclean2016theory} use both classical and quantum resources to solve potentially difficult problems.  This approach is particularly promising for current quantum computers of limited size and power \cite{preskill2018quantum}. Several variants of hybrid quantum algorithms have recently been demonstrated, such as the Variational Quantum Eigensolver (VQE) for quantum chemistry and related applications \cite{kandala2017hardware,peruzzo2014variational,hempel2018quantum,o2016scalable,kokail2019self}, and the Quantum Approximate Optimization Algorithm (QAOA) for graph or other optimization problems \cite{farhi2014quantum,otterbach2017unsupervised}. Hybrid quantum algorithms can also be used for generative models, which aim to learn representations of data in order to make subsequent tasks easier. Applications of generative modeling include computer vision \cite{zhu2017unpaired}, speech synthesis \cite{oord2016wavenet}, the inference of missing text \cite{bowman2015generating}, de-noising of images \cite{bengio2013generalized}, and chemical design \cite{gomez2018automatic}. Here, we apply a hybrid quantum learning scheme on a trapped ion quantum computer \cite{debnath2016demonstration} to accomplish a generative modeling task.

Data-driven quantum circuit learning (DDQCL) is a hybrid framework for generative modeling of classical data where the model consists of a parameterized quantum circuit \cite{benedetti2019generative}. The model is trained by sampling the output of a quantum computer and updating the circuit parameters using a classical optimizer. After convergence, the optimal circuit produces a quantum state that captures the correlations in the training data sets. Hence the trained circuit serves as a generative model for the training data. Theoretical results suggest that such generative models have more expressive power than widely used classical neural networks \cite{du2018expressive,Gaoeaat9004}. This is because instantaneous quantum polynomial circuits -- special cases of the parameterized quantum circuits used for generative modeling -- cannot be efficiently simulated by classical means. 

\begin{figure}[htbp]
\centering
 \includegraphics[width=0.6\textwidth]{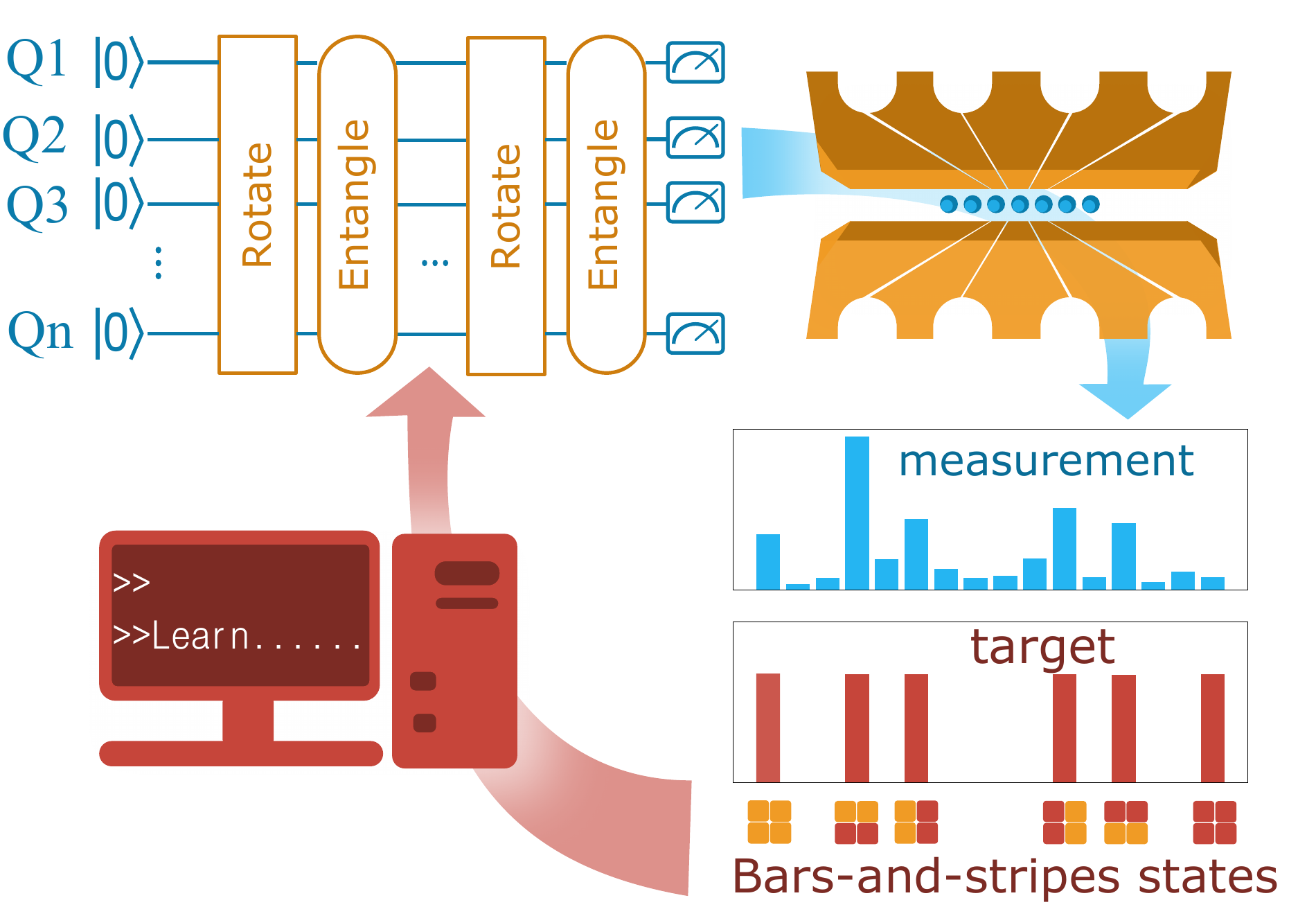}
 \caption{Data-driven quantum circuit learning (DDQCL) is a hybrid quantum algorithm scheme that can be used for generative modeling, illustrated here by the example of 2-by-2 Bars and Stripes (BAS) data. From top left, clockwise: A parametrized circuit is initialized at random. Then at each iteration, the circuit is executed on a trapped ion quantum computer. The probability distribution of measurement is compared on a classical computer against the BAS target data set. Next, the quantified difference is used to optimize the parametrized circuit. This learning process is iterated until convergence.} \label{fig:pipeline}
 \end{figure}

The Bars-and-Stripes (BAS) data set is a canonical body of synthetic data for generative modeling \cite{mackay2003information}. It can be easily visualized in terms of images containing horizontal bars or vertical stripes, where each pixel represents a qubit.  Here, we use the uniformly distributed 2-by-2 BAS shown in Fig.\ref{fig:pipeline} in a proof-of-principle generative modeling task on a trapped-ion quantum computer. This is the first successful demonstration of generative quantum circuits trained on multi-qubit quantum hardware. We note that there has been a single-qubit experiment in this context \cite{hu2019quantum}. We compare the performance of different classical optimization algorithms and conclude that Bayesian optimization shows significant advantages over Particle Swarm Optimization for this task.

The experiment is performed on four qubits within a seven-qubit fully programmable trapped ion quantum computer \cite{landsman2019verified} (see Method). With individual addressing and readout of all qubits, the system can perform sequences of gates from a universal gates set, composed of Ising gates and arbitrary rotations \cite{debnath2016demonstration}. In order to run the large number of variational circuit instances necessary for the data-driven learning, we calibrate single- and two-qubit gates and execute lists of circuits in an automated fashion. 

\begin{figure}[htbp]
\centering
\includegraphics[width=0.6\linewidth]{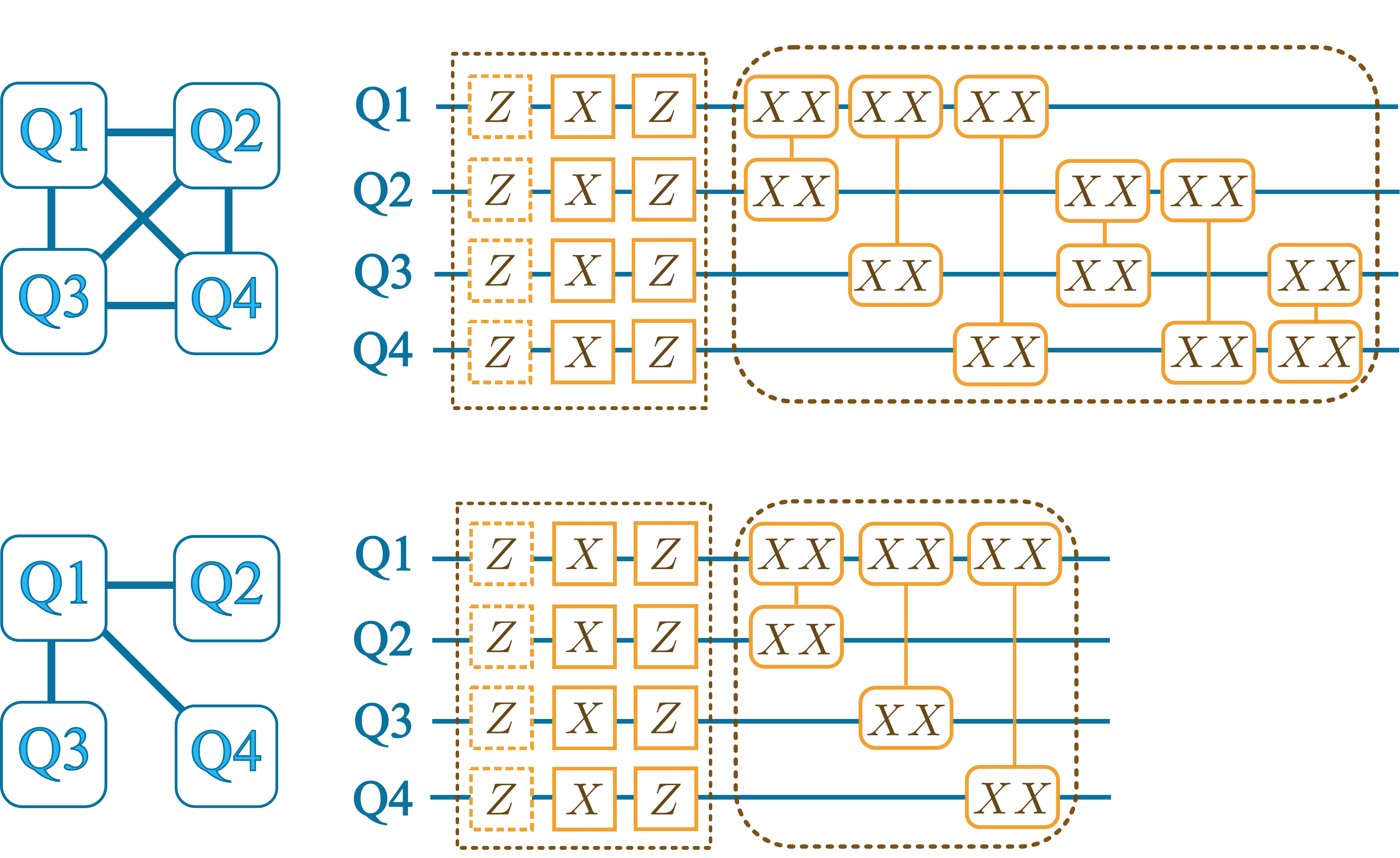}
\caption{Connectivity graphs and corresponding training circuits. Top: Fully-connected training circuit layer, with layers of rotations (square boxes) and entanglement gates (rounded boxes) between any pair of the four qubits.  Bottom: Star-connectivity training circuit layer, with restricted entangling gates. In either case, each rotation (denoted by X or Z) and each entanglement gate (denoted by XX) includes a distinct control parameter, for a total of 18 parameters for the fully-connected circuit layer and 15 parameters for the star-connected circuit layer. We remove the first Z rotation (dashed square box) acting on the initial state $\ket{0}$ , resulting in 14 and 11 parameters, respectively. The connectivity figures on the left define the mapping between the four qubits and the pixels of the BAS images (see Fig.\ref{fig:pipeline}).}
\label{fig:connectivity}.
\end{figure}

The training pipeline is illustrated in Fig. \ref{fig:pipeline}. The quantum circuits are structured as layers of parameterized gates. We use two types of layers, involving single-qubit rotations and two-qubit entangling gates. A single-qubit layer sandwiches an X-rotation between two Z-rotations on each qubit $i$, or $R_{z}^{(i)}(\alpha_{i})R_{x}^{(i)}(\beta_{i})R_{z}^{(i)}(\gamma_{i})$, involving twelve rotation parameters for the four qubits (see Fig. \ref{fig:connectivity}).
An entangling layer applies Ising or XX gates between all pairs of qubits according to any imposed connectivity graph. This is expressed as a sequence of $XX^{i,j}(\chi_{i,j})$ operations as shown in Fig. \ref{fig:connectivity}), with up to six entangling parameters \cite{debnath2016demonstration} for four qubits.  Due to the universality of this gate set, a sufficiently long sequence of layers of these two types can produce arbitrary unitaries. 

At the start of DDQCL, all the rotation and entangling parameters are initialized with random values. Next the circuit is repeatedly executed on the trapped ion quantum computer in order to reconstruct the state distribution. A classical computer then compares the measured distribution with the target distribution and quantifies the difference using a cost function (see Method for details). A classical optimization algorithm then varies the parameters.  We iterate the entire process until convergence.

We impose two distinct connectivity graphs in a four-qubit circuit: all-to-all and star, as shown in Fig.\ref{fig:connectivity}.  With star connectivity, entanglement between certain qubit-pairs cannot occur within a single gate layer, which means more layers are necessary for certain target distributions. Comparing the training process between circuits of different connectivity provides insight into the performance of DDQCL algorithms on platforms with more limited interaction graphs.

For each connectivity graph, we add layers until the goal of reproducing the BAS data with the trained model is achieved. The match between training data and model is limited by noise, experimental throughput rate (how fast the system can process circuits), and sampling errors. The cost function used in optimization scores the result, but a successful training process must be able to generate data that can be qualitatively recognized as a BAS pattern to ensure that the system provides usable results in the spirit of generative modeling in machine learning \cite{theis2015note}.

We now describe the classical optimization strategies for the training algorithm. Although gradient-based approaches were recently proposed for DDQCL\cite{liu2018differentiable}, we employ gradient-free optimization schemes that appear less sensitive to noise and experimental throughput. We explore two such schemes: Particle Swarm Optimization (PSO) \cite{kennedy1995particle} and Bayesian Optimization (BO) \cite{frazier2018tutorial}. PSO is a stochastic optimization scheme commonly used in machine learning that works by creating many ``particles" randomly distributed across parameter space that explore the landscape collaboratively. We limit the number of particles to twice the number of parameters.
BO is a global optimization paradigm that can handle the expensive sampling of many-parameter functions. It works by maintaining a surrogate model of the underlying cost function and, at each iteration, updates the model to guide the search for the global minimum. Essentially, the problem of optimizing the real cost is replaced with that of optimizing the surrogate model, which is designed to be a much easier optimization problem. We use OPTaaS, a BO software package developed by Mind Foundry and adapted for this work.

\section*{Results}

\begin{figure}[htb]
\centering
 \includegraphics[width=\textwidth]{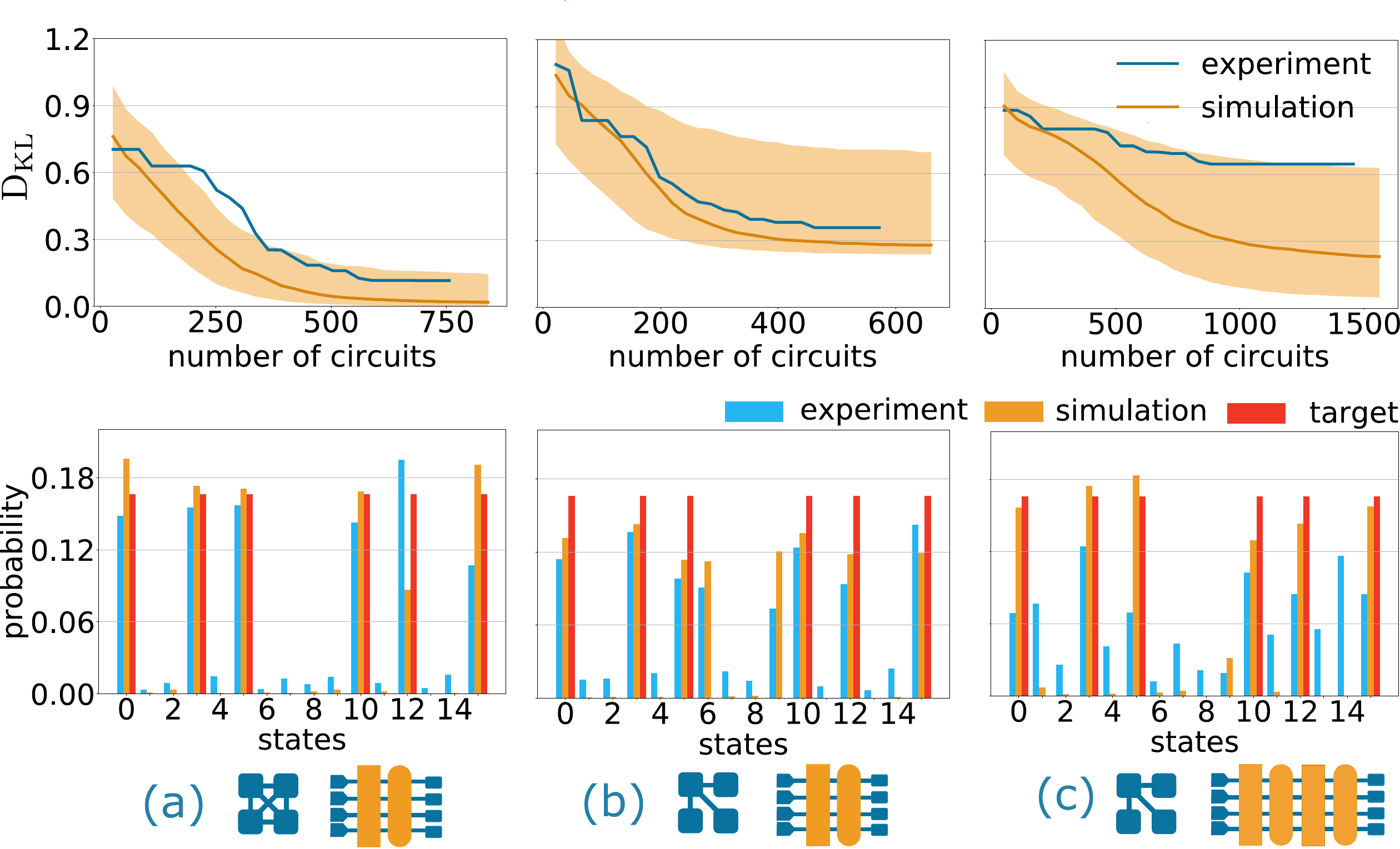}
 \caption{Quantum circuit training results with Particle Swarm optimization (PSO), with simulations (orange) and trapped ion quantum computer results (blue). Column (a) corresponds to a circuit with one layer of single qubit rotations (square boxes) and one layer of entanglement gates (rounded boxes) of all-to-all connectivity.  The circuit converges well to produce the bars-and-stripes (BAS) distribution. Columns (b) and (c) correspond to a circuit with two and four layers and star-connectivity, respectively. In (b), the simulation shows imperfect convergence with two extra state components (6 and 9), due to the limited connectivity, and the experimental results follow the simulation. In (c), the simulation shows convergence to the BAS distribution, but the experiment fails to converge despite performing 1,400 quantum circuits. The optimization is sensitive to the choice of initialization seeds. To illustrate the convergence behavior, the shaded regions span the 5th-95th percentile range of random seeds (500 for (a) and (b), 1000 for (c), and the orange curve shows the median.
 The two-layer circuits have 14 and 11 parameters for (a) all-to-all- and (b) star-connectivity, while the (c) star-connectivity circuit with four layers has 26 parameters. The number of PSO particles used is twice the number of parameters, and each training sample is repeated $5000$ times.  Including circuit compilation, controller-upload time, and classical PSO optimization, each circuit instance takes about 1 min to be processed, in addition to periodic interruptions for the re-calibration of gates.}\label{fig:pso_result}
\end{figure}

Results from PSO optimization are shown in Fig. \ref{fig:pso_result}. We first simulate the training procedure using a classical simulator in place of the quantum processor (orange plots in Fig. \ref{fig:pso_result}). Since the PSO method is sensitive to the initial "seed" values of the particles, we simulate the convergence for many different random seeds (see Fig.\ref{fig:pso_result}). We choose a seed that converges quickly and reliably under simulated sampling error to start the training procedure on the trapped ion quantum computer illustrated in Fig.\ref{fig:pipeline}. We iterate the training until it converges (blue plots in Fig.\ref{fig:pso_result}). In practice, which seeds are successful is unknown, and different seeds need to be tried experimentally until a good model is obtained. This incurs an additional cost in the form of multiple independent DDQCL training rounds.

For all-to-all connectivity, we find that a circuit with one rotation gate layer and one entangling gate layer is able to produce the desired BAS distribution (Fig. \ref{fig:pso_result}a). This is not the case for the star-connected circuit, with the closest state having two additional components in the superposition (states 6 and 9 in Fig. \ref{fig:pso_result}b). With two additional layers, the star-connected circuit is able to model the BAS distribution (orange plots of Fig. \ref{fig:pso_result}c). In the experiment however (blue plots in Fig. \ref{fig:pso_result}c), the PSO is unable to converge to an acceptable solution even using the best pre-screened seed value and sufficient sample statistics. We conclude that PSO fails because the throughput rate is too low for effectively training the circuit in the face of gate imperfections.

\begin{figure}[htbp]
\centering
 \includegraphics[width=\textwidth]{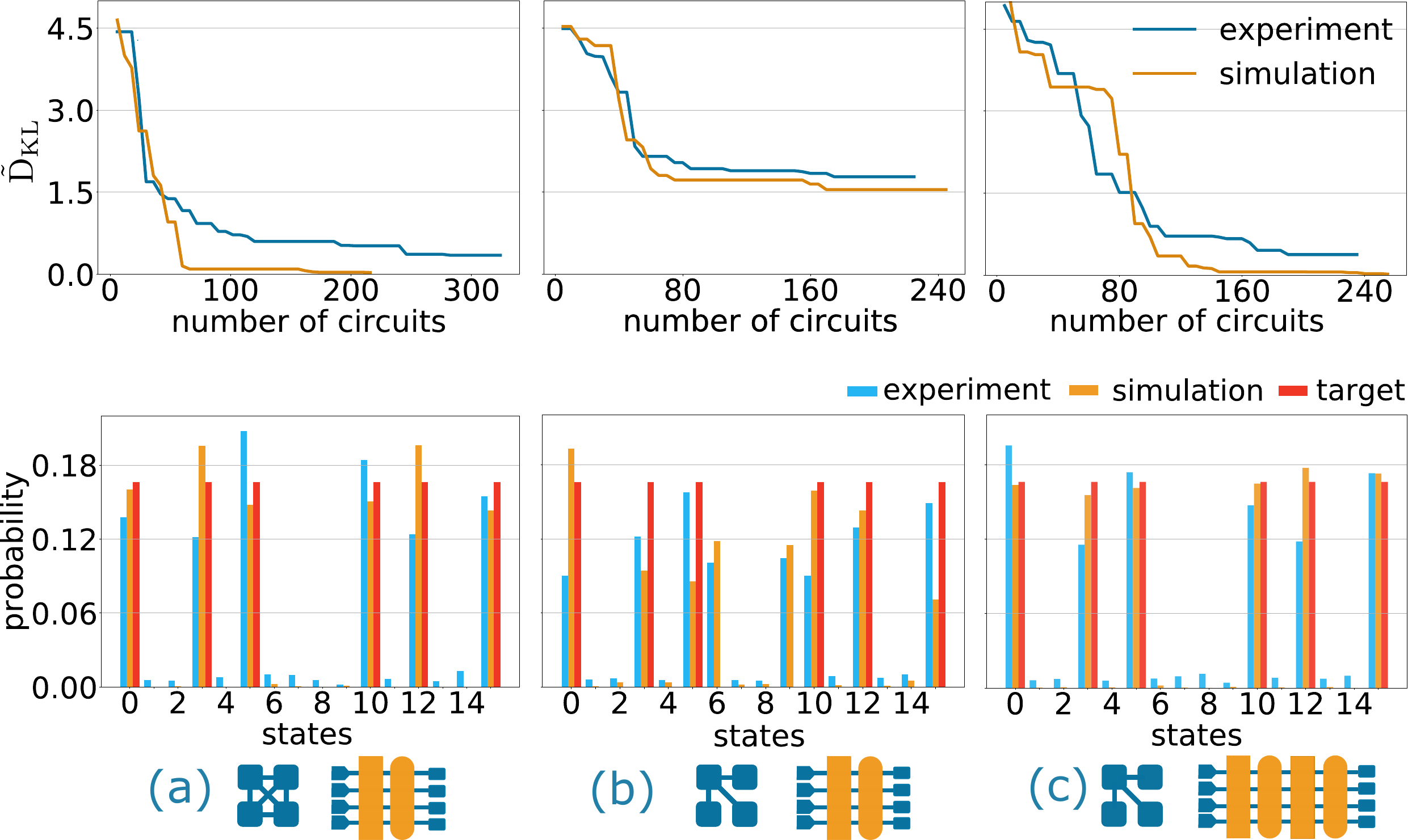}
 \caption{Quantum circuit training results with Bayesian optimization (BO), with simulations (orange) and trapped ion quantum computer results (blue). Column (a) corresponds to a circuit with two layers of gates and all-to-all connectivity.  Columns (b) and (c) correspond to a circuit with two and four layers and star-connectivity, respectively. Convergence is much faster than with PSO (Fig. \ref{fig:pso_result}). Unlike the PSO results, the four-layer star-connected circuit in (c) is trained successfully, and no prior knowledge enters BO process. As before, the two-layer circuits have 14 and 11 parameters for (a) all-to-all- and (b) star-connectivity, while the (c) star-connectivity circuit with four layers has 26 parameters. We use a batch of 5 circuits per iteration, and each training sample is repeated $5000$ times.  Including circuit compilation, controller-upload time, and BO classical optimization, each circuit instance takes 2-5 minutes, depending on the amount of accumulated data.} \label{fig:optaas_result}
\end{figure}

For these reasons, we instead employ a Bayesian optimization scheme for the circuit training procedure. We find that all circuits experimentally converge in agreement with the simulations, as shown in Fig. \ref{fig:optaas_result}.  Moreover, even the star-connected circuit with four layers now produces a recognizable BAS distribution (Fig. \ref{fig:optaas_result}c). In contrast to PSO, BO dramatically reduces the number of samples needed for training and does not require any pre-selection of random seeds or other prior knowledge of the cost-function landscape.

BO updates the surrogate model using the experimental result of every iteration.  Therefore, the classical part of each BO iteration consumes more time than with PSO, where the time cost on the classical optimizer is negligible. However, the BO procedure converges faster to the desired BAS distribution. More generally, these examples highlight the need to balance quantum and classical resources in order to produce acceptable performance and run time in a hybrid quantum algorithm.

As a measure of the performance of the various training procedures, we compute the Kullback-Leibler (KL) divergence \cite {kullback1951information} and the qBAS score (an alternative performance measure suggested in \cite{benedetti2019generative}) of the experimental results at the end of each DDQCL training run, shown in Table~\ref{table:main}.
We also compute the entanglement entropy (S) averaged over all two plus two qubit partitions assuming a pure state~\cite{higuchi2000entangled}, estimated via simulation of the quantum state from the trained circuits. The entanglement entropy quantifies the level of entanglement of a state, thus indicates how difficult it is to produce such state. This metric shows that the successfully trained circuits generate states that are consistent with a high level of entanglement.  As a reference, the entanglement entropy of a GHZ state over any partition is $S=1$.

\clearpage

\begin{table}[htb]
\begin{center}
\begin{tabular}{ |l|c|c|c|c| } 
\hline
\mbox{ }circuits\hspace{29mm} &\mbox{ }optimizer\mbox{ } &\mbox{ } $D_{\rm{KL}}$ \mbox{ } &\mbox{}qBAS score\mbox{} &\mbox{ }$S$\mbox{ }\\
\hline
\multirow{2}{4em}{\mbox{ }\includegraphics[height=9mm]{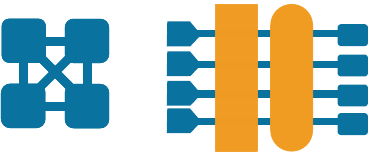}} & PSO & 0.116 &0.91&1.628 \\ 
& BO & 0.094&0.91&1.659 \\
\hline
\multirow{2}{4em}{\mbox{ }\includegraphics[height=9mm]{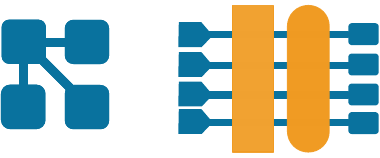}} & PSO & 0.357 &0.74&0.9950 \\ 
& BO & 0.328&0.77&0.9999 \\ 
\hline
\multirow{2}{4em}{\mbox{ }\includegraphics[height=9mm]{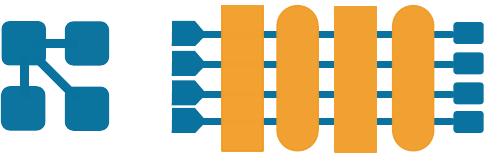}} & PSO & 0.646 &0.59&0.8867 \\ 
& BO & 0.100&0.91&1.709 \\ 
\hline
\end{tabular}
\end{center}
\caption{KL divergence ($D_{KL}$, see Materials and Methods), qBAS score, and entanglement entropy ($S$) for the state obtained at the end of each of the DDQCL training on hardware, for various circuits and classical optimizers used.} \label{table:main}
\end{table}

\section*{Discussion}

This demonstration of generative modeling using reconfigurable quantum circuits of up to 26 parameters represents one of the most powerful hybrid quantum applications to date. With ongoing engineering improvements\cite{wright2019benchmarking}, we expect the system to grow in both qubit number and gate quality. This approach can be scaled up to handle larger data sets with increased qubit number by adapting the cost function for sparser sampling \cite{benedetti2019generative}. Moreover, this procedure can be adapted for other types of hybrid quantum algorithms.

Classical optimization techniques for hybrid quantum algorithms on intermediate-scale quantum computer do not always succeed \cite{hamilton2018generative}. Recent work suggests that typical cost functions for medium to large scale variational quantum circuits landscape resemble ``barren plateaus" \cite{mcclean2018barren}, making optimization hard. As quantum computers scale up for larger problems, the cost of classical optimization such as BO must be weighed against the quantum algorithmic advantage.  
%Thus, advances in both quantum computer hardware and classical optimizing strategies will underlie continued progress in hybrid quantum-classical algorithms.

\section*{Materials and Methods}

\subsection*{Trapped Ion Quantum Computer}
The trapped ion quantum computer used for this study consists of a chain of seven single $^{171}$Yb$^+$ ions confined in a Paul trap and laser cooled close to their motional ground state. Each ion provides one physical qubit in the form of a pair of states in the hyperfine-split $^2S_{1/2}$ ground level with an energy difference of 12.642821 GHz, which is insensitive to magnetic fields to first order. The qubits are collectively initialized into $\ket{0}$ through optical pumping, and state readout is accomplished by state-dependent fluorescence detection \cite{olmschenk07}. Qubit operations are realized via pairs of Raman beams, derived from a single 355-nm mode-locked laser \cite{debnath2016demonstration}. These optical controllers consist of an array of individual addressing beams and a counter-propagating global beam that illuminates the entire chain. Single qubit gates are realized by driving resonant Rabi rotations of defined phase, amplitude, and duration. Single-qubit rotations about the z-axis, are performed by classically advancing/regarding the phase of the optical beatnote applied to the particular qubit. Two-qubit gates are achieved by illuminating two selected ions with beat-note frequencies near motional sidebands and creating an effective Ising spin-spin interaction via transient entanglement between the two qubits and the motion in the trap\cite{molmer99,solano99,milburn00}. Since our particular scheme involves multiple modes of motion, we use an amplitude modulation scheme to disentangle the qubit state from the motional state at the end of the interaction \cite{choi2014optimal}. Typical single-qubit gate fidelities are $99.5(2)\%$. Typical two-qubit gate fidelities are $98-99\%$, with fidelity mainly limited by  residual entanglement of the qubit states to the motional state of the ions, coherent crosstalk and driving intensity noise from classical imperfections in our optical controllers.

In our experiment, the effect of the gate errors is seen as an offset in the cost function after convergence. An improvement in gate fidelity will reduce this offset. But the convergence behavior of an ideal system (as shown in the simulations in Fig.\ref{fig:pso_result} and Fig.\ref{fig:optaas_result}) is not significantly faster than the actual experimental system. This is because it is limited by the classical optimization routine.

The trapped ion quantum architecture is scalable to a much larger number of qubits, as atomic clock qubits are perfectly replicable and do not suffer idle errors (T1 and T2 times are essentially infinite).  All of the errors in scaling arise from the classical controllers, such as applied noise on the trap electrodes and laser beam intensity fluctuations. Fundamental errors (like spontaneous scattering from the control laser beams) are not expected to play a role until our gates approach $99.99\%$ fidelity. However, as the qubit number grows beyond about 20-30, we expect to sacrifice full connectivity, as gates will only be performed with high fidelity between any qubit and its 15-20 nearest neighbors. 

Another limitation is the sampling rate on the quantum computer. This is limited by technical issues on the current experiment, and can be improved, e.g. by increasing the upload speed of the experimental control system.

\subsection*{Classical Optimizers: PSO and BO}

We explore two different classical optimizer in this study: Particle Swarm Optimization(PSO) and Bayesian Optimization(BO).

PSO is a gradient-free optimization method inspired by the social behaviour of some animals. Each particle represents a candidate solution and moves within the solution space according to its current performance and the performance of the swarm. Three hyper-parameters control the dynamics of the swarm: a cognition coefficient $c_1$, a social coefficient $c_2$, and an inertia coefficient $w$ \cite{kennedy1995particle}.

Concretely, each particle consists of a position vector $\theta_i$ and a velocity vector $v_i$. At iteration $t$ of the algorithm, the velocity of particle $i$ for the coordinate $d$ is updated as

\begin{equation}
v_{i,d}^{(t+1)} = w  v_{i,d}^{(t)} + c_1  r_{1,d}^{(t)}  ( p_{i,d}^{(t)} - \theta_{i,d}^{(t)} ) + c_2 r_{2,d}^{(t)} (g_d^{(t)} - \theta_{i,d}^{(t)}) ,
\label{e:velocity}
\end{equation}

where $r_{1,d}^{(t)}$ and $r_{2,d}^{(t)}$ are random numbers sampled from the uniform distribution in [0,1] for every dimension and every iteration, $p_i^{(t)}$ is the particle's best position, $g^{(t)}$ is the swarm's best position. The position is then updated as

\begin{equation}
\theta_i^{(t+1)} = \theta_i^{(t)} + v_i^{(t)} ,
\label{e:position}
\end{equation}

In our problem, each particle corresponds to a point in parameter space of the quantum circuit. For example, in the fully connected circuit with two layers, each particle consists of an instance of the 14 parameters. Recall, however, that parameters are angles and are therefore periodic; We customized the PSO updates above to use this information. In Eq.~\eqref{e:velocity}, $p_{i,d}^{(t)}$ and $\theta_{i,d}^{(t)}$ can be thought of as two points on a circle. Instead of using the standard displacement $p_{i,d}^{(t)} - \theta_{i,d}^{(t)}$, we use the angular displacement, that is the signed length of the minor arc on the unit circle. We use the same definition of displacement for the swarm's best position $g_{i,d}^{(t)}$. Finally, in Eq.~\eqref{e:position}, we make sure to express angles always using their principal values.

In our experiments, we set the number of particles to twice the number of parameters of the circuit. Position and velocity vectors of each particle are initialized from the uniform distribution. For the coefficients we use $c_1=c_2=1$ and $w=0.5$.

Bayesian Optimisation is a powerful global optimisation paradigm. It is best suited to finding optima of multi-modal objective functions that are expensive to evaluate. There are two main features that characterize the a BO process: the surrogate model and an acquisition function. 

The surrogate model is non-parametric model of the objective function. At each iteration, the surrogate model is updated using the sampled points in parameter space. The package used in this study is OPTaaS by MindFoundry. It implements the surrogate model as regression using Gaussian Process\cite{rasmussen2003gaussian}. A kernel (or correlation function) characterizes the Gaussian process, we use a Matern 5/2 as it provides the most flexibility. 

The acquisition function is computed from the surrogate model. It is used to select points for evaluation during the optimization. It trades off exploration against exploitation. The acquisition function of a point has a high value if the cost function is expected to give a significant improvement over historically sampled points, or if the uncertainty of the point is high, according to the surrogate model. A simple and well known acquisition function, Expected Improvement\cite{brochu2010tutorial}, is employed here.  

In our case, OPTaaS also leverages the cyclic symmetry of the angles by embedding the parameter space into a metric space with the appropriate topology, effectively allowing the Gaussian Process surrogate model to be placed over a hyper-torus, rather than a hyper-cube. This greatly alleviates the so-called curse of dimensionality\cite{bellman2015adaptive}, and allows for much more efficient use of samples of the objective function.

It is key in Bayesian Optimisation to adequately optimise the acquisition function during each iteration. OPTaaS puts considerable computational resources towards this non-convex optimisation problem.

There are two major reasons why the BO out performs PSO in our specific case. First, PSO spends significant amount of computation resource exploring trajectories far from optimal, while BO mitigates it by the use of acquisition function. Second, the maintenance of the surrogate model enable us to make much better use of the information from the historical exploration of the parameter space. 

\subsection*{Cost Functions}
We use a cost function to quantify the difference between the target BAS distribution and the experimental measurements of the circuit. The cost functions used to implement the training are variants of the original Kullback-Leibler Divergence ($D_{KL}$) \cite{kullback1951information}:
\begin{equation}\label{eq:KLD}
D_{KL}(p,q)=-\sum_{i}p(i)\log\frac{q(i)}{p(i)}.
\end{equation}

Here $p$ and $q$ are two distributions.

$D_{KL}(p,q)$ is an information theoretic measure of how two probability distribution differ. If base 2 for the logarithm is used, it quantifies the expected number of extra bits required to store samples from p when an optimal code designed for q is used instead. It can be shown that $D_{KL}(p,q)$ is non-negative, and is zero if and only if p=q. However, it is asymmetric in the arguments and does not satisfy the triangle inequality. Therefore $D_{KL}(p,q)$ is not a metric.

The KL divergence is a very general measure, but it is not always well-defined, e.g. if an element of the domain is supported by $p$ and not by $q$, the measure will diverge. This problem may occur quite often if $D_{KL}(p,q)$ is estimated from samples and if the dimensionality of the domain is large. For PSO, we use the clipped negative log-likelihood cost function \cite{benedetti2019generative},
\begin{equation}\label{eq:Negative_LL}
C_{nll}=-\sum_{i}p(i)\log\{\max[\epsilon ,q(i)]\}.
\end{equation}
Here we set $p$ as the target distribution. Thus Eq.\ref{eq:Negative_LL} is equivalent to Eq.\ref{eq:KLD} up to a constant offset, so the optimization of these two functions is equivalent. $\epsilon$ is a small number (0.0001 here) used to avoid a numerical singularity when $q(i)$ is measured to be zero.

For BO, we use the clipped symmetrized Kullback-Leibler (KL) divergence as the cost function 
\begin{equation}\label{eq:symmetrized_KL}
\tilde{D}_{KL}(p,q)=D_{KL}[\max(\epsilon,p),
\max(\epsilon,q)]+D_{KL}[\max(\epsilon,q),\max(\epsilon,p)].
\end{equation}
This is found to be the most reliable variant of $D_{KL}$ for BO.

%\bibliography{scibib}

\begin{thebibliography}{10}

\bibitem{mcclean2016theory}
J.~R. McClean, J.~Romero, R.~Babbush, A.~Aspuru-Guzik, The theory of
  variational hybrid quantum-classical algorithms.
\newblock {\it New Journal of Physics\/} {\bf 18}, 023023 (2016).

\bibitem{preskill2018quantum}
J.~Preskill, Quantum {C}omputing in the {NISQ} era and beyond.
\newblock {\it {Quantum}\/} {\bf 2}, 79 (2018).

\bibitem{kandala2017hardware}
A.~Kandala, A.~Mezzacapo, K.~Temme, M.~Takita, M.~Brink, J.~M. Chow, J.~M.
  Gambetta, Hardware-efficient variational quantum eigensolver for small
  molecules and quantum magnets.
\newblock {\it Nature\/} {\bf 549}, 242 (2017).

\bibitem{peruzzo2014variational}
A.~Peruzzo, J.~McClean, P.~Shadbolt, M.-H. Yung, X.-Q. Zhou, P.~J. Love,
  A.~Aspuru-Guzik, J.~L. {O}'{B}rien, A variational eigenvalue solver on a
  photonic quantum processor.
\newblock {\it Nature communications\/} {\bf 5}, 4213 (2014).

\bibitem{hempel2018quantum}
C.~Hempel, C.~Maier, J.~Romero, J.~McClean, T.~Monz, H.~Shen, P.~Jurcevic,
  B.~P. Lanyon, P.~Love, R.~Babbush, A.~Aspuru-Guzik, R.~Blatt, C.~F. Roos,
  Quantum chemistry calculations on a trapped-ion quantum simulator.
\newblock {\it Phys. Rev. X\/} {\bf 8}, 031022 (2018).

\bibitem{o2016scalable}
P.~Oâ€™Malley, R.~Babbush, I.~Kivlichan, J.~Romero, J.~McClean, R.~Barends,
  J.~Kelly, P.~Roushan, A.~Tranter, N.~Ding, {\it et~al.\/}, Scalable quantum
  simulation of molecular energies.
\newblock {\it Physical Review X\/} {\bf 6}, 031007 (2016).

\bibitem{kokail2019self}
C.~Kokail, C.~Maier, R.~van Bijnen, T.~Brydges, M.~Joshi, P.~Jurcevic,
  C.~Muschik, P.~Silvi, R.~Blatt, C.~Roos, {\it et~al.\/}, Self-verifying
  variational quantum simulation of lattice models.
\newblock {\it Nature\/} {\bf 569}, 355 (2019).

\bibitem{farhi2014quantum}
E.~Farhi, J.~Goldstone, S.~Gutmann, A quantum approximate optimization
  algorithm.
\newblock {\it MIT-CTP/4610\/}  (2014).

\bibitem{otterbach2017unsupervised}
J.~Otterbach, R.~Manenti, N.~Alidoust, A.~Bestwick, M.~Block, B.~Bloom,
  S.~Caldwell, N.~Didier, E.~S. Fried, S.~Hong, {\it et~al.\/}, Unsupervised
  machine learning on a hybrid quantum computer.
\newblock {\it arXiv preprint arXiv:1712.05771\/}  (2017).

\bibitem{zhu2017unpaired}
J.-Y. Zhu, T.~Park, P.~Isola, A.~A. Efros, {\it Proceedings of the IEEE
  international conference on computer vision\/} (2017), pp. 2223--2232.

\bibitem{oord2016wavenet}
A.~Van Den~Oord, S.~Dieleman, H.~Zen, K.~Simonyan, O.~Vinyals, A.~Graves,
  N.~Kalchbrenner, A.~Senior, K.~Kavukcuoglu, Wavenet: A generative model for
  raw audio.
\newblock {\it CoRR abs/1609.03499\/}  (2016).

\bibitem{bowman2015generating}
S.~R. Bowman, L.~Vilnis, O.~Vinyals, A.~M. Dai, R.~Jozefowicz, S.~Bengio,
  Generating sentences from a continuous space.
\newblock {\it SIGNLL Conference on Computational Natural Language Learning
  (CONLL), 2016\/}  (2016).

\bibitem{bengio2013generalized}
Y.~Bengio, L.~Yao, G.~Alain, P.~Vincent, {\it Advances in Neural Information
  Processing Systems\/} (2013), pp. 899--907.

\bibitem{gomez2018automatic}
R.~G{\'o}mez-Bombarelli, J.~N. Wei, D.~Duvenaud, J.~M. Hern{\'a}ndez-Lobato,
  B.~S{\'a}nchez-Lengeling, D.~Sheberla, J.~Aguilera-Iparraguirre, T.~D.
  Hirzel, R.~P. Adams, A.~Aspuru-Guzik, Automatic chemical design using a
  data-driven continuous representation of molecules.
\newblock {\it ACS central science\/} {\bf 4}, 268--276 (2018).

\bibitem{debnath2016demonstration}
S.~Debnath, N.~M. Linke, C.~Figgatt, K.~A. Landsman, K.~Wright, C.~Monroe,
  Demonstration of a small programmable quantum computer with atomic qubits.
\newblock {\it Nature\/} {\bf 536}, 63 (2016).

\bibitem{benedetti2019generative}
M.~Benedetti, D.~Garcia-Pintos, O.~Perdomo, V.~Leyton-Ortega, Y.~Nam,
  A.~Perdomo-Ortiz, A generative modeling approach for benchmarking and
  training shallow quantum circuits.
\newblock {\it npj Quantum Information\/} {\bf 5}, 45 (2019).

\bibitem{du2018expressive}
Y.~Du, M.-H. Hsieh, T.~Liu, D.~Tao, The expressive power of parameterized
  quantum circuits.
\newblock {\it arXiv preprint arXiv:1810.11922\/}  (2018).

\bibitem{Gaoeaat9004}
X.~Gao, Z.-Y. Zhang, L.-M. Duan, A quantum machine learning algorithm based on
  generative models.
\newblock {\it Science Advances\/} {\bf 4} (2018).

\bibitem{mackay2003information}
D.~J. MacKay, D.~J. Mac~Kay, {\it Information theory, inference and learning
  algorithms\/} (Cambridge university press, 2003).

\bibitem{hu2019quantum}
L.~Hu, S.-H. Wu, W.~Cai, Y.~Ma, X.~Mu, Y.~Xu, H.~Wang, Y.~Song, D.-L. Deng,
  C.-L. Zou, {\it et~al.\/}, Quantum generative adversarial learning in a
  superconducting quantum circuit.
\newblock {\it Science advances\/} {\bf 5}, eaav2761 (2019).

\bibitem{landsman2019verified}
K.~A. Landsman, C.~Figgatt, T.~Schuster, N.~M. Linke, B.~Yoshida, N.~Y. Yao,
  C.~Monroe, Verified quantum information scrambling.
\newblock {\it Nature\/} {\bf 567}, 61 (2019).

\bibitem{theis2015note}
L.~Theis, A.~v.~d. Oord, M.~Bethge, A note on the evaluation of generative
  models.
\newblock {\it arXiv preprint arXiv:1511.01844\/}  (2015).

\bibitem{liu2018differentiable}
J.-G. Liu, L.~Wang, Differentiable learning of quantum circuit born machines.
\newblock {\it Physical Review A\/} {\bf 98}, 062324 (2018).

\bibitem{kennedy1995particle}
J.~Kennedy, R.~Eberhart, {\it Proc. IEEE International Conference on Neural
  Networks, Perth, Australia\/} (1995), pp. 1942--1948.

\bibitem{frazier2018tutorial}
P.~I. Frazier, A tutorial on bayesian optimization.
\newblock {\it arXiv preprint arXiv:1807.02811\/}  (2018).

\bibitem{kullback1951information}
S.~Kullback, R.~A. Leibler, On information and sufficiency.
\newblock {\it The annals of mathematical statistics\/} {\bf 22}, 79--86
  (1951).

\bibitem{higuchi2000entangled}
A.~Higuchi, A.~Sudbery, How entangled can two couples get?
\newblock {\it Physics Letters A\/} {\bf 273}, 213--217 (2000).

\bibitem{wright2019benchmarking}
K.~Wright, K.~Beck, S.~Debnath, J.~Amini, Y.~Nam, N.~Grzesiak, J.-S. Chen,
  N.~Pisenti, M.~Chmielewski, C.~Collins, {\it et~al.\/}, Benchmarking an
  11-qubit quantum computer.
\newblock {\it arXiv preprint arXiv:1903.08181\/}  (2019).

\bibitem{hamilton2018generative}
K.~E. Hamilton, E.~F. Dumitrescu, R.~C. Pooser, Generative model benchmarks for
  superconducting qubits.
\newblock {\it arXiv preprint arXiv:1811.09905\/}  (2018).

\bibitem{mcclean2018barren}
J.~R. McClean, S.~Boixo, V.~N. Smelyanskiy, R.~Babbush, H.~Neven, Barren
  plateaus in quantum neural network training landscapes.
\newblock {\it Nature communications\/} {\bf 9}, 4812 (2018).

\bibitem{olmschenk07}
S.~Olmschenk, K.~C. Younge, D.~L. Moehring, D.~N. Matsukevich, P.~Maunz,
  C.~Monroe, Manipulation and detection of a trapped ${\mathrm{yb}}^{+}$
  hyperfine qubit.
\newblock {\it Phys. Rev. A\/} {\bf 76}, 052314 (2007).

\bibitem{molmer99}
K.~M\o{}lmer, A.~S\o{}rensen, Multiparticle entanglement of hot trapped ions.
\newblock {\it Phys. Rev. Lett.\/} {\bf 82}, 1835--1838 (1999).

\bibitem{solano99}
E.~Solano, R.~L. de~Matos~Filho, N.~Zagury, Deterministic bell states and
  measurement of the motional state of two trapped ions.
\newblock {\it Phys. Rev. A\/} {\bf 59}, R2539--R2543 (1999).

\bibitem{milburn00}
G.~Milburn, S.~Schneider, D.~James, Ion trap quantum computing with warm ions.
\newblock {\it Fortschritte der Physik\/} {\bf 48}, 801--810 (2000).

\bibitem{choi2014optimal}
T.~Choi, S.~Debnath, T.~A. Manning, C.~Figgatt, Z.-X. Gong, L.-M. Duan,
  C.~Monroe, Optimal quantum control of multimode couplings between trapped ion
  qubits for scalable entanglement.
\newblock {\it Phys. Rev. Lett.\/} {\bf 112}, 190502 (2014).

\bibitem{rasmussen2003gaussian}
C.~E. Rasmussen, {\it Summer School on Machine Learning\/} (Springer, 2003),
  pp. 63--71.

\bibitem{brochu2010tutorial}
E.~Brochu, V.~M. Cora, N.~De~Freitas, A tutorial on bayesian optimization of
  expensive cost functions, with application to active user modeling and
  hierarchical reinforcement learning.
\newblock {\it arXiv preprint arXiv:1012.2599\/}  (2010).

\bibitem{bellman2015adaptive}
R.~E. Bellman, {\it Adaptive control processes: a guided tour\/}, vol. 2045
  (Princeton university press, 2015).

\end{thebibliography}

\bibliographystyle{Science}

\section*{Acknowledgments}
We thank C. Figgatt for helpful discussion. 
This work was supported by the ARO with funds from the Intelligence Advanced Research Projects Activity (IARPA) LogiQ program (Grant Number W911NF16-1-0082), the Army Research Office (ARO) MURI program on Modular Quantum Circuits (Grant Number W911NF1610349), the AFOSR MURI program on Optimal Quantum Measurements (Grant Number 5710003628), the NSF STAQ Practical Fully-Connected Quantum Computer Project, and the NSF Physics Frontier Center at JQI (Grant Number PHY0822671). L. Egan is additionally funded by NSF award DMR-1747426.
\subsection*{Authors' contributions}
D. Z, N. M. L, M. B, K. A. L, A. P and C. M designed the research. D. Z, N. M. L, M. B, K. A. L, N. H. N, C. H. A, A. P, L. E, and O. P collected and analyzed data. D. Z, M. B, A. P, N. K, A. G and C. B contributed to the software used in this study. All authors contributed to this manuscript.
\subsection*{Competing interests}
C.M. is a founding scientist of IonQ, Inc. All other authors declare that they have no competing interests.
\subsection*{Data availability}
%The data presented in the figures and those supporting the other findings in this study are available from the corresponding author upon request.
All data needed to evaluate the conclusions in the paper are present in the paper and/or the Supplementary Materials. Additional data related to this paper may be requested from the corresponding author upon request.

%Here you should list the contents of your Supplementary Materials -- below is an example. 
%You should include a list of Supplementary figures, Tables, and any references that appear only in the SM. 
%Note that the reference numbering continues from the main text to the SM.
% In the example below, Refs. 4-10 were cited only in the SM. 

\clearpage

 % For your review copy (i.e., the file you initially send in for
% evaluation), you can use the {figure} environment and the
% \includegraphics command to stream your figures into the text, placing
% all figures at the end.  For the final, revised manuscript for
% acceptance and production, however, PostScript or other graphics
% should not be streamed into your compliled file.  Instead, set
% captions as simple paragraphs (with a \noindent tag), setting them
% off from the rest of the text with a \clearpage as shown  below, and
% submit figures as separate files according to the Art Department's
% instructions.

\end{document}